\documentclass[iop]{emulateapj}
\usepackage[T1]{fontenc}
\usepackage{microtype,amsmath}
\usepackage[colorlinks,urlcolor=blue,citecolor=blue,linkcolor=blue]{hyperref}

\slugcomment{Submitted ApJL 2013 Jan 10; revised 2013 Mar 12; accepted 2013
  Mar 13}
\shorttitle{Quasi-Quiescent Radio Emission from 2M\,1047}
\shortauthors{Williams \textit{et al.}}

\newcommand\citeeg[1]{\citep[e.g.,][]{#1}}
\newcommand\ujy{\ensuremath{\mu\text{Jy}}}
\newcommand\ujybm{$\mu$Jy~bm$^{-1}$}
\newcommand\vtwom{2M\,1047+21} 
\newcommand\twom{\object{\vtwom}}
\newcommand\ftwom{\object{2MASS\,J10475385+2124234} (\vtwom\ hereafter)}
\newcommand\vjthirt{2M\,1315$-$26}
\newcommand\jthirt{\object{\vjthirt}}
\newcommand\fjthirt{\object{2MASS\,J13153094-2649513AB} (\jthirt\ hereafter)}
\newcommand\apx{\ensuremath{\sim}}
\newcommand\uv{$u$\mbox{-}$v$}
\newcommand\fpow[3]{\left(\frac{#1}{#2}\right)^{#3}}
\newcommand\ha{H$\alpha$}

\begin{document}

\title{Quasi-Quiescent Radio Emission from the First Radio-Emitting T Dwarf}
\author{
  Peter K.~G. Williams\altaffilmark{1},
  Edo Berger\altaffilmark{1},
  and
  B. Ashley Zauderer\altaffilmark{1}
}
\email{pwilliams@cfa.harvard.edu}
\altaffiltext{1}{Harvard-Smithsonian Center for Astrophysics, 60 Garden Street,
  Cambridge, MA 02138, USA}

\begin{abstract}
  Radio detections of ultracool dwarfs provide insight into their magnetic
  fields and the dynamos that maintain them, especially at the very bottom of
  the main sequence, where other activity indicators dramatically weaken.
  Until recently, radio emission was only detected in the M and L dwarf
  regimes, but this has changed with the Arecibo detection of rapid polarized
  flares from the T6.5 dwarf 2MASS\,J10475385+2124234. Here, we report the
  detection of quasi-quiescent radio emission from this source at 5.8~GHz
  using the Karl G. Jansky Very Large Array. The spectral luminosity is $L_\nu
  = (2.2 \pm 0.7) \times 10^{12}$~erg~s$^{-1}$~Hz$^{-1}$, a factor of \apx100
  times fainter than the Arecibo flares. Our detection is the
  lowest-luminosity yet achieved for an ultracool dwarf. Although the emission
  is fully consistent with being steady, unpolarized, and broadband, we find
  tantalizing hints for variability. We exclude the presence of short-duration
  flares as seen by Arecibo, although this is not unexpected given estimates
  of the duty cycle. Follow-up observations of this object will offer the
  potential to constrain its rotation period, electron density, and the
  strength and configuration of the magnetic field. Equally important,
  follow-up will address the question of whether the electron cyclotron maser
  instability, which is thought to produce the flares seen by Arecibo, also
  operates in the very different parameter regime of the emission we detect,
  or whether instead this ultracool dwarf exhibits both maser and
  gyrosynchrotron radiation, potentially originating from substantially
  different locations.
\end{abstract}


\defcitealias{rw12}{RW12}
\newcommand\rw{\citetalias{rw12}}

\section{Introduction}
\label{s.intro}

Although there is now abundant evidence that very low-mass stars and brown
dwarfs (collectively, ultracool dwarfs: UCDs) can host significant magnetic
activity, the origins and detailed nature of this activity are not
well-understood. While there is consensus that the $\alpha\Omega$ dynamo that
powers the solar magnetic field cannot operate in these fully-convective
objects, modeling of turbulent dynamos yields a wide
range of predictions for the strength, morphology, and time-dependence of UCD
fields \citeeg{ddyr93,ck06,dsb06,b08}. These poorly-constrained yet basic
physical properties affect the internal structure, atmospheric conditions, and
dynamics of low-mass stars, brown dwarfs, and giant planets.

The evidence for UCD magnetic activity comes from a variety of tracers.
Standard indicators such as X-ray and \ha\ emission \citeeg{rbmb00,whw+04}
have been augmented by Zeeman-Doppler imaging
\citep[ZDI;][]{thezdi,dmp+08,mdp+08} and FeH spectroscopy \citep{rb06,rb07}. A
particularly noteworthy discovery was the detection of UCD radio emission
\citep{bbb+01}, which is much stronger than would be expected from stellar
scaling relations \citep{gb93,bg94}. Radio emission is an increasingly
important observable because other tracers become much more difficult to
exploit for the very latest-type dwarfs, in which rapid rotation obliterates
Zeeman signals, the overall optical luminosity decreases greatly, and X-ray
and \ha\ emission fall off precipitously \citep{smf+06,bbf+10,gmr+00,whw+04}.
This last effect is possibly due to increasingly neutral atmospheres
\citep{mbs+02}, enhanced trapping of energetic electrons, centrifugal
stripping of the corona, or other effects decreasing the efficiency with which
the coronal plasma is heated \citep{bbf+10}.

A deeper understanding of the convective dynamo will likely be built upon a
clearer picture of the magnetic phenomenology of UCDs across the widest
possible range of temperatures and Rossby numbers. Measurements of activity in
the latest dwarfs, which are extreme in both these parameters, are thus vital.
Until recently, the latest-type UCD to be detected in the radio was the L3.5
object \object{2MASS\,J00361617+1821104} \citep{b02,brr+05,had+08}. However,
in the past year, two later-type sources have been detected. The L5e+T7 binary
\fjthirt, known to have strong long-lived \ha\ emission \citep{h02b,bsg+11},
was shown by \citet{bmzb13} to have quiescent radio emission. Even more
surprisingly, the T-dwarf barrier was broken by \citet[RW12 hereafter]{rw12},
who detected circularly-polarized radio bursts from the T6.5 dwarf
\ftwom\ with the Arecibo observatory. At a distance of 10.6~pc and a
temperature of a mere \apx 900~K \citep{vhl+04}, \twom\ represents a new
benchmark in the study of cool, magnetically-active objects.

In this Letter, we report the detection of \twom\ with the Karl G. Jansky Very
Large Array (VLA) with properties significantly different from those of the
flares detected by \rw\ (\S\ref{s.obs}). We discuss its implications as well
as some potential paths forward (\S\ref{s.disc}).

\section{Observations \& Analysis}
\label{s.obs}

\twom\ was spectroscopically identified as a T~dwarf by \citet{bkb+99}. It was
later assigned spectral types of T7 \citep[optical;][]{bklb03} and T6.5
\citep[near infrared;][]{bgl+06}. \citet{bklb03} also made a marginal
(2.2$\sigma$) detection of \ha\ emission, measuring a flux of
$f_{\text{H}\alpha} = (5.9 \pm 2.7) \times 10^{-18}$~erg~cm$^{-2}$~s$^{-1}$.
Astrometric monitoring revealed a distance of $10.6 \pm 0.4$~pc and a proper
motion of $1728.4 \pm 7.7$~mas~yr$^{-1}$ \citep{vhl+04}. \citet{vhl+04} also
computed a bolometric luminosity of $\ell_\text{bol} = -5.36$ for \twom, where
$\ell = \log_{10}(L/L_\sun)$. Combining these results, $\ell_{\text{H}\alpha}
- \ell_\text{bol} = -5.3 \pm 0.2$, well above the trend for typical UCDs
\citep{bbf+10}. There is no evidence for \twom\ being a binary system, with
\citet{bkr+03} ruling out companions at separations $\gtrsim$4~AU with mass
ratios $\gtrsim$0.4 and \citet{cmp+11} finding weaker constraints. Before the
radio detection by \rw, \citet{b06b} reported a flux density upper limit of
45~\ujy\ at 8.46~GHz ($\ell_r - \ell_\text{bol} < -5.52$, where $L_r = \nu
L_\nu$)\footnote{We note that the limit relative to bolometric quoted in
  \citet{mbr12} is incorrect.}.

Using the astrometric parameters of \citet{vhl+04}, the predicted location of
\twom\ at the time of our VLA observations is RA = 10:47:52.03, Dec =
+21:24:16.7 (FK5~J2000). We use a simple Monte Carlo approach to estimate the
1$\sigma$-confidence ellipse in this position to have major and minor axes of
0.19$''$ and 0.09$''$ at a position angle of 77\degr. This is much smaller
than the \apx 1$''$ astrometric uncertainties of our observation in the VLA C
configuration at \apx6~GHz and much larger than the difference between the FK5
and ICRS frames.

We observed \twom\ in a two-hour session starting on 2012~April~17.02 UT
(project 12A-451). The total correlated bandwidth is 2048~MHz, with two
basebands centered at 5000 and 6750~MHz and divided into 512 channels each.
Standard calibration observations were obtained, using \object{3C\,286} as the
flux density standard and bandpass calibrator and the quasar
\object{J1051+2119} (0.9\degr\ from \twom) serving as the complex antenna gain
calibrator. The total integration time on \twom\ is 80~minutes. The data were
calibrated using standard procedures in the CASA software system
\citep{thecasa}. Radio frequency interference was flagged manually.

\begin{figure}[p]
\plotone{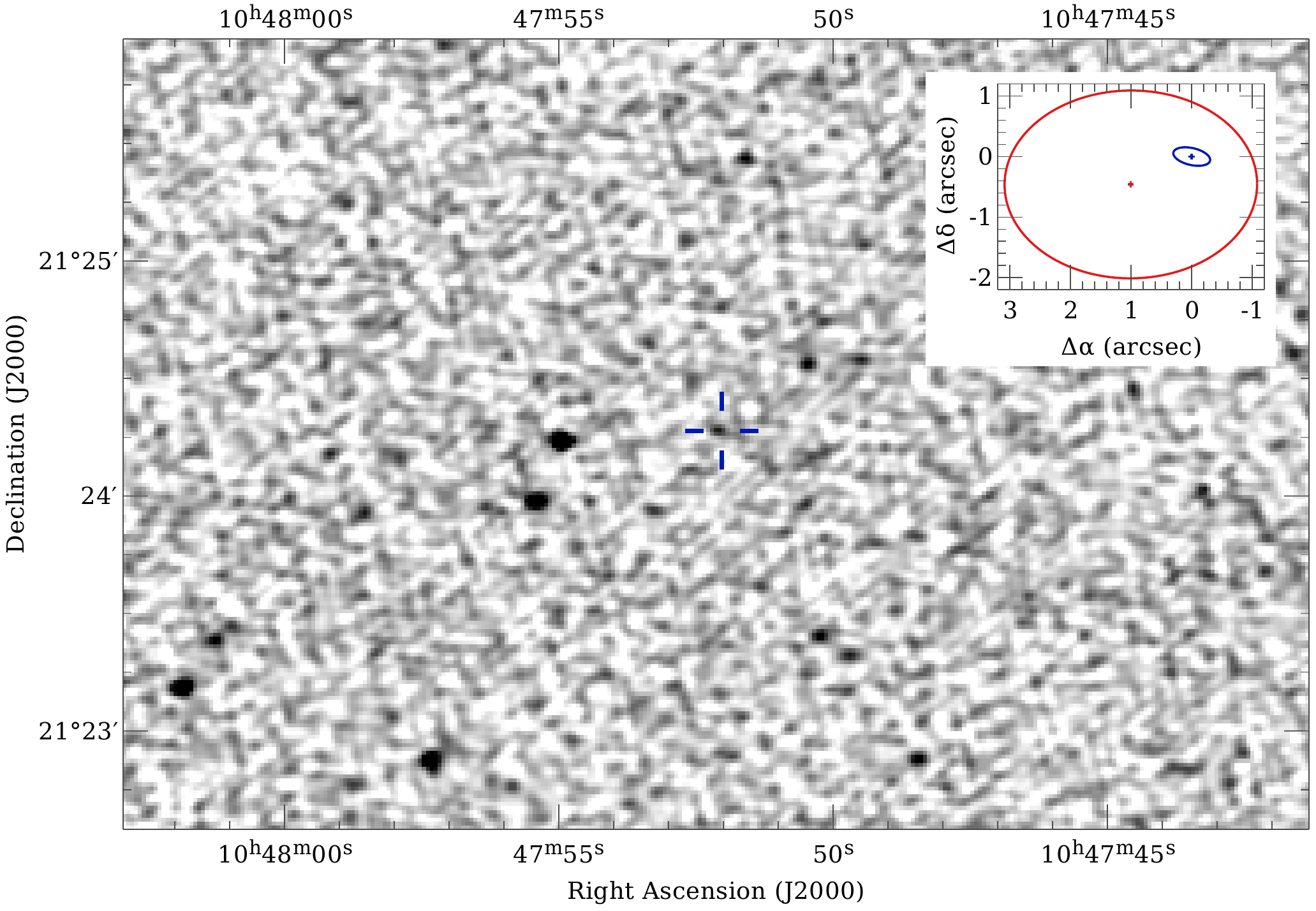}
\caption{VLA detection of \vtwom. The grayscale is linear black-to-white from
  -5 to +20 \ujy. The background rms is 3.6 \ujy. The crosshairs in the main
  image are centered on the predicted location of \vtwom\ from \citet{vhl+04},
  which is independent of our radio observations. The enlarged inset compares
  the 1$\sigma$ confidence regions of the VLA (red ellipse) and \citet[blue
    ellipse]{vhl+04} positions.}
\label{f.vlaimg}
\end{figure}

We created a deep Stokes~I image of 2048$\times$2048 pixels, each
1$''$$\times$1$''$, using multi-frequency synthesis \citep{themfs} and CASA's
multifrequency CLEAN algorithm with two spectral Taylor series terms for each
CLEAN component; this approach models both the flux and spectral index of each
source. The synthesized beam is $4.2'' \times 3.6''$ and the reference
frequency is 5.84~GHz, with data taken at frequencies between 4.51 and
7.24~GHz contributing to the imaging process. In the resulting image of the
initial Taylor term (i.e., brightness at the reference frequency) we detect a
4.6$\sigma$ (peak-to-rms) unresolved source at RA = 10:47:52.10, Dec =
+21:24:16.2 (ICRS~J2000) with 1$\sigma$ positional uncertainties of 0.9$''$
and 0.6$''$ in RA and declination, respectively. The separation from the IR
astrometric prediction is 1.0$''$ at a position angle of 117\degr. Taking into
account only the uncertainties on the VLA source position, and not the
uncertainties in the IR position or the astrometric calibration of the VLA
image, the separation is $1.3\sigma$. The relevant portion of the image is
shown in Figure~\ref{f.vlaimg}. Because we search for a source at only one
\textit{a priori} image position, the stringent detection criteria that have
been advocated for blind radio source searches \citeeg{fko+12} do not apply.

Fitting the image region near the source with a Gaussian shaped like the
synthesized beam, we find a flux density of $16.5 \pm 5.1$~\ujy\ while the rms
residual of the deconvolution process in the neighborhood is 3.6~\ujybm, well
above the \apx0.2~\ujybm\ confusion limit for this observing setup
\citep{c02}. The uncertainty in the flux density is that reported by the
fitting process, which is not generically equal to the background rms due to
the two positional parameters. The overall significance of the peak, however,
is set by the rms. We identify this source with \twom. The likelihood of a
spurious association is small: \citet{fwkk91} found the density of radio
sources brighter than 16~\ujy\ at 5~GHz to be
\apx$2\times10^{-4}$~arcsec$^{-2}$, making the probability of background
source contamination \apx$10^{-3}$ (assigning our search area to be the
synthesized beam size). We counted peaks >16~\ujy\ in the central region of
our image and found a density of \apx $2.4\times10^{-4}$~arcsec$^{-2}$,
agreeing well with the \citet{fwkk91} value. Our measurement does not account
for primary beam attenuation, but the correction for that effect is a factor
of a few, not large enough to change the conclusion of a low probability of
chance coincidence.

The deconvolution process also determines a nominal spectral index of $\alpha
= 0.9 \pm 1.0$ ($S_\nu \propto \nu^\alpha$). We also imaged the two VLA
basebands separately with only one spectral Taylor term, i.e., assuming
spectrally flat emission, and detected emission consistent with the position
of \twom\ in both images. The best-fit flux density of the source in the
lower-frequency image is $18.0\pm9.2$~\ujy, larger than that in the
higher-frequency image, $16.7\pm9.7$~\ujy; in both cases, the best-fit
position is \apx0.5 of a beam away from the best-fit position in the combined
image. These fluxes tentatively suggest $\alpha < 0$, but the uncertainties
are large. Combining our results with the upper limit of \citet{b06b} at
8.46~GHz implies $\alpha \lesssim 2.7$ if variability is not a factor. Without
calibrating the instrumental polarization leakages, we imaged the Stokes~V
parameter and found no emission at the location of \twom. We estimate $|V|/I
\lesssim 80\%$ ($3\sigma$).

The radio emission detected by \rw\ comprised flares with flux densities up to
\apx 2~mJy on timescales of minutes. Thanks to the excellent sensitivity of
the VLA, less than a second of integration time would be needed for a
5$\sigma$ detection of these flares in our data, so superb time resolution of
any flares should be possible. We searched for flares and other variability
using a visibility-domain analysis. We subtracted the emission of all other
detectable sources from the \uv\ data and summed the residual visibilities
after rephasing to the position of \twom. Ignoring noise and calibration
errors, the real part of the summed rephased visibility is the source flux,
while the imaginary part is zero. Given that noise and calibration errors are
always present, the variation of the imaginary part provides an assessment of
the uncertainty in the real part. The mean real part is $16.5$~\ujy\ and the
variance associated with the imaginary parts
$\sigma_\text{im}/\sqrt{N_\text{im}} = 4.6$~\ujy, agreeing with the
image-domain analysis. Our calibrated data have a 5-second cadence (reduced
from the 1-second cadence of the raw data), so any Arecibo-like flares should
be apparent as several consecutive \apx$12\sigma$ bins. We find no variability
of the kind observed by \rw, with only two bins of $>$$3\sigma$ significance,
which is consistent with the expectation from Gaussian noise of 2.6 such
events.

We Hanning-smoothed the data over a variety of timescales to search for
phenomena of longer duration but lower significance. Figure~\ref{f.photom}
shows the data smoothed on 30~s and 6.5~min timescales. There is a
rise-and-fall pattern in the 6.5-minute cadence data suggestive of a
\apx60-\ujy\ enhancement over a 20-min timescale. The data are however fully
consistent with a constant flux density, matching this model with a reduced
$\chi^2$ of 1.05. We imaged a 36-minute span around the event ($56034.041 \le
\text{MJD[TT]} \le 56034.066$) and found a source of flux density
$37.2\pm6.8$~\ujy\ with a background rms of 4.8~\ujy, yielding a 7.8$\sigma$
detection. The significance of the difference between this value and the mean
flux density is lessened, however, because we searched a range of start times
and durations precisely to maximize this difference. The possible flare or
enhancement seen in the binned data is a factor of \apx30 times less luminous
and evolves on a timescale \apx10 times longer than the \rw\ flares. The total
energy of the potential event cannot be compared to that of the \rw\ flares
without knowledge of its spectrum; if it is a broader-band phenomenon (see
below), the two event classes could have approximately equal energies.

\section{Discussion and Conclusions}
\label{s.disc}

\begin{figure}[p]
\plotone{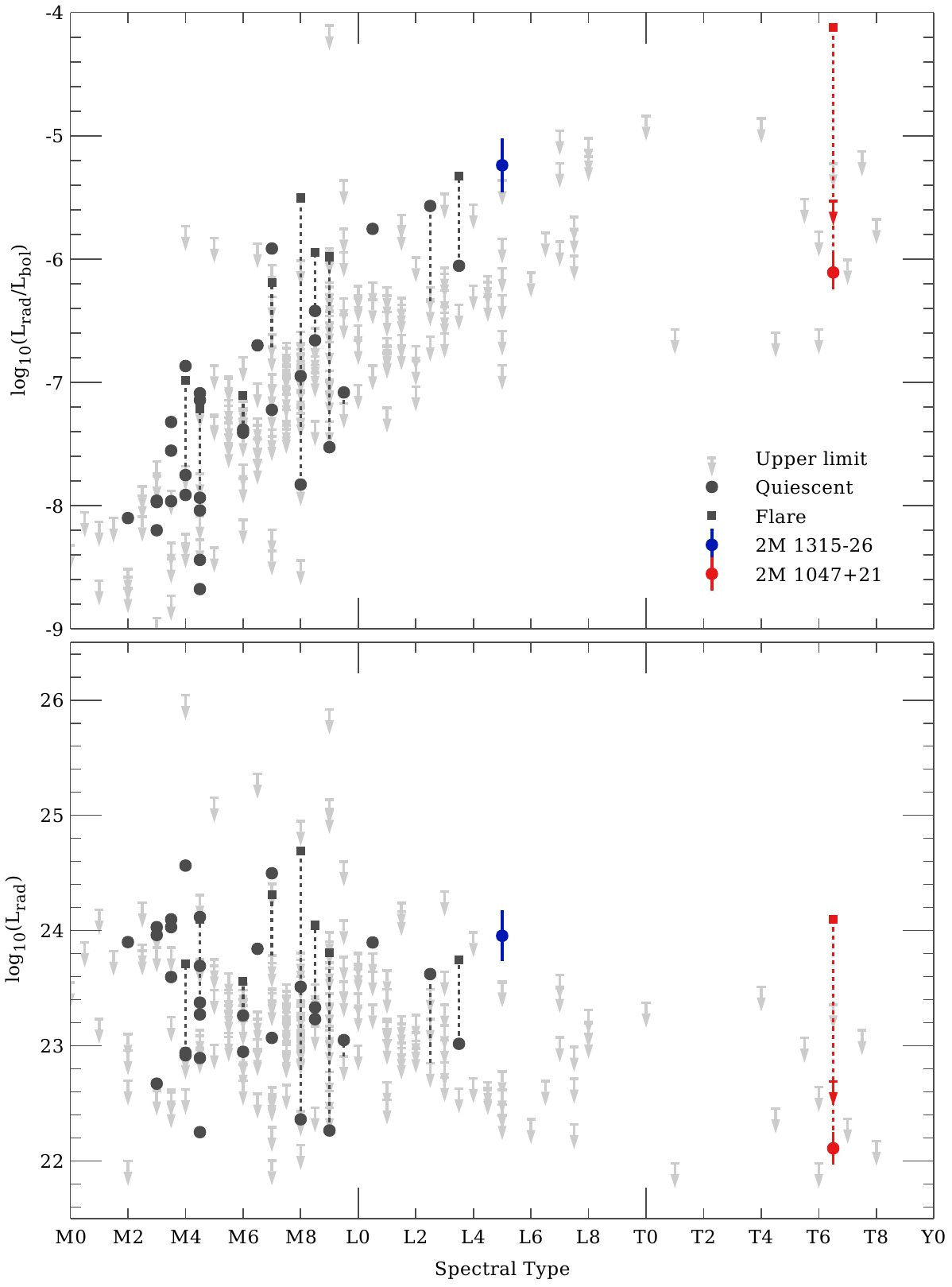}
\caption{Ultracool dwarf radio luminosity as a function of spectral type.
  \textit{Upper panel}: normalized to the bolometric luminosity. \textit{Lower
    panel}: unnormalized. The background data are from \citet[and references
    therein]{mbr12}. The measurement of \vjthirt\ by \citet{bmzb13} is in
  \textit{blue}. Recent measurements of \vtwom\ are in \textit{red} (flare:
  \rw, upper limit: \citet{b06b}, quiescent: this work, assuming steady
  emission). It is the only radio-detected ultracool dwarf of spectral type T,
  and its flares as detected by \rw\ are an order of magnitude more luminous,
  relative to $L_\text{bol}$, than those from other UCDs. Our detection is the
  lowest-luminosity yet achieved for a UCD.}
\label{f.rlrad}
\end{figure}

Assuming that the emission from \twom\ is indeed steady and using the distance
reported by \citet{vhl+04}, the radio spectral luminosity is $L_\nu = (2.2 \pm
0.7) \times 10^{12}$~erg~s$^{-1}$~Hz$^{-1}$ or $\ell_\text{rad} -
\ell_\text{bol} = -6.11 \pm 0.14$. We compare this value to
previously-detected UCDs in Figure~\ref{f.rlrad}. Not only is \twom\ by far
the coolest UCD to be detected in the radio, but it also has flares that are
two orders of magnitude brighter than its quiescent emission, second only to
the M8 dwarf \object{DENIS J1048.0-3956} \citep{bp05} in the observed dynamic
range of its luminosity. If the emission that we detect is not truly
quiescent, the dynamic range must be even larger.

If the radio emission of \twom\ is indeed quiescent and unpolarized, it is
suggestive of a gyrosynchrotron process with a brightness temperature
\begin{align}
T_b &= \frac{c^2}{2 k_B \nu^2} \frac{S_\nu}{\Omega} \cr
 &\approx \left(10^{6.5}\text{ K}\right) \fpow{\nu}{\text{GHz}}{-2}
   \fpow{S_\nu}{\ujy}{} \fpow{d}{\text{pc}}{2} \fpow{r}{R_J}{-2} \cr
 &\approx \left(10^8\text{ K}\right) \fpow{r}{R_J}{-2},
\end{align}
where $r$ is the effective radius of the emitting region and the final line
plugs in the relevant values from our observation. Length scales of $r
\lesssim 0.01 R_J$ would be problematic for this interpretation due to the
brightness temperature constraints of gyrosynchrotron emission \citep{d85}.
Gyrosynchrotron emission is broadband with a spectral peak, the location of
which depends on local physical parameters. Further characterization of the
radio spectrum, ideally paired with measurements in other bands, could yield
conditional estimates of the magnetic field strength and electron number
density \citeeg{b06b}.

Another candidate emission mechanism is the electron cyclotron maser
instability \citep[ECMI;][]{theecm,t06}. Stereotypical ECMI emission is highly
polarized, narrowband, and time-variable, making it a natural explanation for
the bursts observed by \rw. Its applicability to the emission we detect is
less clear. The emission that we detect shows no evidence of strong
polarization, is broadband, and may be variable but clearly does not show the
strong spikes seen in \twom\ by \rw\ or \object{2MASS~J07464256+2000321} by
\citet{brpb+09}. \citet{had+06,had+08} have argued that continuous particle
acceleration in a region of varying magnetic field strength can generate ECMI
emission that is broad in frequency and time, and that propagation effects can
depolarize it, potentially yielding results that are indistinguishable from
gyrosynchrotron. Their arguments are partially motivated by the detection of
rapid variability in the unpolarized emission of \object{2MASS
  J00361617+1821104} that would violate gyrosynchrotron brightness temperature
constraints given the implied size of the emission region. If the emission we
detect is really concentrated in one or possibly two weak flares of 10--20~min
duration, then the same brightness-temperature argument might apply to \twom.
We note that \citet{had+08} also argue that cooler objects should tend toward
more polarized emission, as propagation effects are less significant in their
increasingly neutral atmospheres; if the emission we detect is strongly
depolarized, it would challenge the ECMI interpretation. On the other hand,
the marginal \ha\ emission \citep{bklb03} may suggest that the atmosphere of
\twom\ may not be fully neutral.

\rw\ detect three radio flares in \apx26~hr of observing for a rate of
\apx0.1~hr$^{-1}$. If the flares occur at random times, this is consistent
with the lack of events in our 2~hr observation. It is also possible that the
flares occur periodically, as observed in a significant fraction of
radio-active UCDs \citeeg{brr+05,brpb+09,hbl+07,had+08}, but with only three
detections this hypothesis cannot be tested meaningfully. The best candidate
periodicity for the three \rw\ events is nominally $2.08156 \pm 0.00003$~hr
(M.~Route, priv. comm.), slightly longer than the length of time for which
\twom\ is observable with Arecibo. The time baseline between the Arecibo
observations and ours is sufficiently long that, even taking this periodicity
and its uncertainty at face value, the phasing of our observations relative to
the Arecibo flares is virtually unconstrained. We thus cannot shed any light
on the regularity of the flares.

\begin{figure}[p]
\plotone{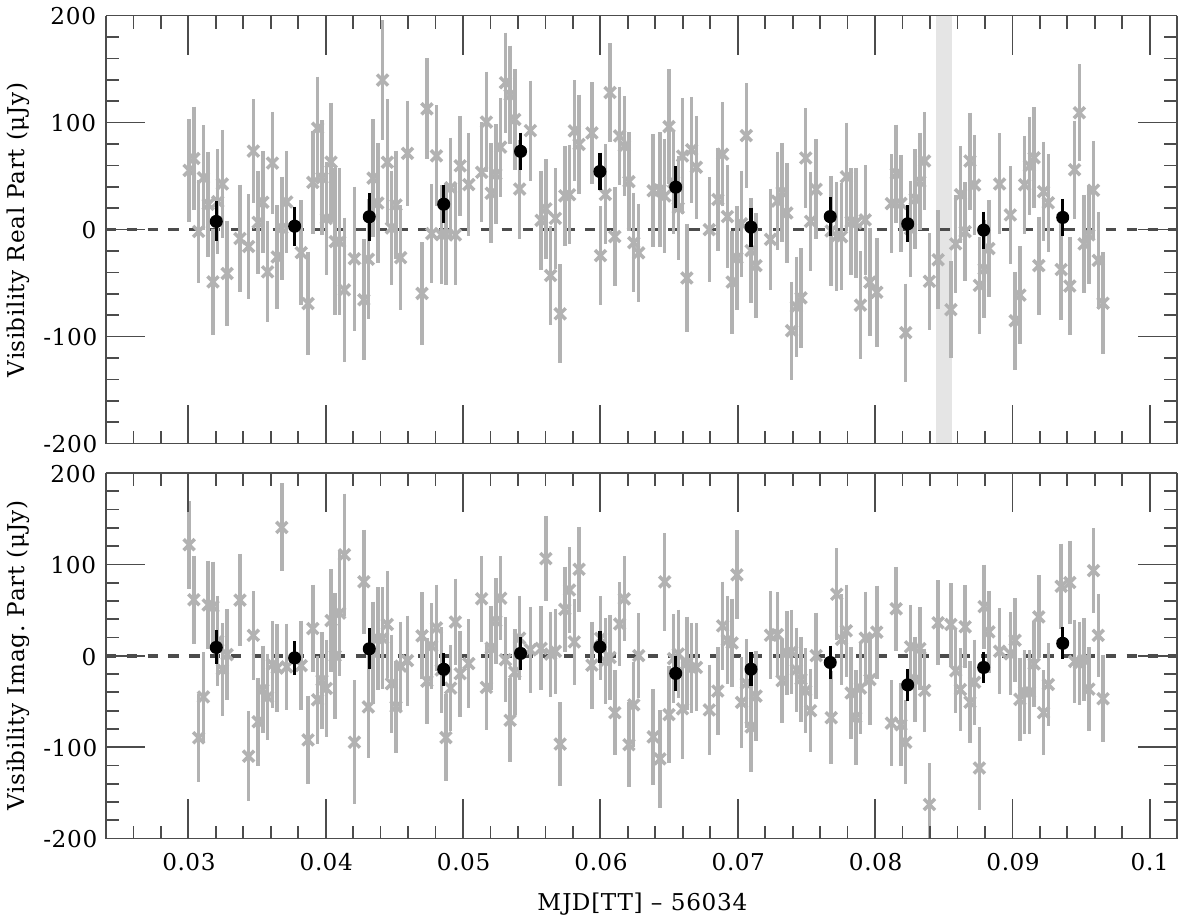}
\caption{Visibility-domain photometry of \vtwom\ as described in
  \S\ref{s.obs}. \textit{Upper panel}: the real parts of the summed
  visibilities, giving the nominal source flux density. \textit{Lower panel}:
  the imaginary parts, which would be zero if there were no noise or
  calibration errors. \textit{Light exes}: the summed visibilities after
  Hanning smoothing on a 1-minute timescale and decimating by a factor of 6,
  yielding a 30-second cadence in which adjacent measurements are weakly
  correlated. \textit{Dark circles}: the summed visibilities smoothed on a
  6.5-minute timescale and decimated by a factor of 78, so that adjacent
  measurements are uncorrelated. The light curve is suggestive of variability
  with a timescale of \apx0.5~hr and a peak flux density of \apx60~\ujy. The
  width of the \textit{shaded band} is 100~s, the timescale of the flares
  detected by \rw; its horizontal position is arbitrary. At flux densities of
  \apx 2~mJy, such flares would register as \apx ten $12\sigma$ bins in our
  5-second-cadence data (approximately accounting for the flare rise and fall)
  or \apx two $30\sigma$ bins in the 30-second-cadence data plotted here.}
\label{f.photom}
\end{figure}

Additional observations of this unique object are clearly needed. The most
pressing question to resolve is whether the quasi-quiescent emission we detect
is truly steady or whether it consists of very small flares as suggested by
the binned data in Figure~\ref{f.photom}. Fortunately, the sensitivity of the
fully-upgraded VLA makes this a relatively straightforward proposition to
test. Longer integrations allow for a more rigorous characterization of the
variability of this faint source, constraining both the quasi-quiescent
component as well as the duty cycle of the bright flares detected by \rw. A
detection of periodicity in either of these could probe the rotation rate of
\twom, which is currently unknown. If the quasi-quiescent emission is
genuinely steady, it would imply that the magnetic field is likewise
long-lived and potentially stable on a timescale of \apx years. If the
quasi-quiescent emission is variable, the relationship between its modulation
and the temporal behavior of the \rw\ flares could shed light on the magnetic
field topology of \twom.

Further observations to pin down the radio spectrum and polarization of
\twom\ are also desirable. A measurement of the magnetic field strength
associated with the quasi-quiescent emission could be compared to the 1.7~kG
estimate from the flares of \rw\ to provide more clues as to the configuration
of the radio-emitting regions. Strict limits on the polarization fraction
would require a more careful consideration of the level of ionization in the
atmosphere of \twom\ and its relation to the ECMI propagation effects proposed
by \citet{had+08}. Simultaneous multiwavelength observations \citep[radio,
  \ha, optical broadband, X-rays; e.g.][]{bgg+08,bbf+10} could probe the
relationship between the photosphere, chromosphere, and corona, through
monitoring of either a flare event or modulation of the quasi-quiescent
emission.

Both \twom\ and \jthirt\ \citep{bmzb13} are exciting objects that may yield a
wealth of new insights into the mechanisms of radio emission from UCDs. Both
are also prime examples of the discoveries made possible by start-of-the-art,
wide-bandwidth radio telescopes. The VLA in particular promises to be a key
tool for the radio detection of UCDs well below the bottom of the main
sequence. The tenfold increase of its sensitivity over the previous
capabilities enabled the unprecedently low-luminosity detection presented in
this Letter and will make possible the first solid constraints on the UCD
radio luminosity function, successful surveys of targeted subsets of the UCD
population, and detailed studies of exotic dwarfs that are rare and hence
distant.

\acknowledgments

We thank Matthew Route and Aleksander Wolszczan for insight into the Arecibo
results and useful discussion on the manuscript, and the anonymous referee for
comments that improved the paper. We thank the NRAO for granting the
Director's Discretionary Time used to make the observations presented in this
Letter. We acknowledge support for this work from the National Science
Foundation through Grant AST-1008361. The VLA is operated by the National
Radio Astronomy Observatory, a facility of the National Science Foundation
operated under cooperative agreement by Associated Universities, Inc. This
research has made use of NASA's Astrophysics Data System and the SIMBAD
database, operated at CDS, Strasbourg, France.

Facilities: \facility{Karl G. Jansky VLA}

\bibliographystyle{yahapj}
\bibliography{pkgw}{}

\begin{thebibliography}{46}
\expandafter\ifx\csname natexlab\endcsname\relax\def\natexlab#1{#1}\fi

\bibitem[{{Benz} \& {G\"{u}del}(1994)}]{bg94}
{Benz}, A.~O., \& {G\"{u}del}, M. 1994,
  \href{http://adsabs.harvard.edu/abs/1994A\%26A...285..621B}{A\&A, 285, 621}

\bibitem[{{Berger}(2002)}]{b02}
{Berger}, E. 2002, \href{http://dx.doi.org/10.1086/340301}{ApJ, 572, 503}

\bibitem[{{Berger}(2006)}]{b06b}
---. 2006, \href{http://dx.doi.org/10.1086/505787}{ApJ, 648, 629}

\bibitem[{{Berger} {et~al.}(2001){Berger}, {Ball}, {Becker}, {Clarke}, {Frail},
  {Fukuda}, {Hoffman}, {Mellon}, {Momjian}, {Murphy}, {Teng}, {Woodruff},
  {Zauderer}, \& {Zavala}}]{bbb+01}
{Berger}, E., {Ball}, S., {Becker}, K.~M., {et~al.} 2001,
  \href{http://dx.doi.org/10.1038/35066514}{Nature, 410, 338}

\bibitem[{{Berger} {et~al.}(2005){Berger}, {Rutledge}, {Reid}, {Bildsten},
  {Gizis}, {Liebert}, {Mart\'{\i}n}, {Basri}, {Jayawardhana}, {Brandeker},
  {Fleming}, {Johns-Krull}, {Giampapa}, {Hawley}, \& {Schmitt}}]{brr+05}
{Berger}, E., {Rutledge}, R.~E., {Reid}, I.~N., {et~al.} 2005,
  \href{http://dx.doi.org/10.1086/430343}{ApJ, 627, 960}

\bibitem[{{Berger} {et~al.}(2008){Berger}, {Gizis}, {Giampapa}, {Rutledge},
  {Liebert}, {Mart\'{\i}n}, {Basri}, {Fleming}, {Johns-Krull}, {Phan-Bao}, \&
  {Sherry}}]{bgg+08}
{Berger}, E., {Gizis}, J.~E., {Giampapa}, M.~S., {et~al.} 2008,
  \href{http://dx.doi.org/10.1086/524769}{ApJ, 673, 1080}

\bibitem[{{Berger} {et~al.}(2009){Berger}, {Rutledge}, {Phan-Bao}, {Basri},
  {Giampapa}, {Gizis}, {Liebert}, {Mart\'{\i}n}, \& {Fleming}}]{brpb+09}
{Berger}, E., {Rutledge}, R.~E., {Phan-Bao}, N., {et~al.} 2009,
  \href{http://dx.doi.org/10.1088/0004-637x/695/1/310}{ApJ, 695, 310}

\bibitem[{{Berger} {et~al.}(2010){Berger}, {Basri}, {Fleming}, {Giampapa},
  {Gizis}, {Liebert}, {Mart\'{\i}n}, {Phan-Bao}, \& {Rutledge}}]{bbf+10}
{Berger}, E., {Basri}, G., {Fleming}, T.~A., {et~al.} 2010,
  \href{http://dx.doi.org/10.1088/0004-637x/709/1/332}{ApJ, 709, 332}

\bibitem[{{Browning}(2008)}]{b08}
{Browning}, M.~K. 2008, \href{http://dx.doi.org/10.1086/527432}{ApJ, 676, 1262}

\bibitem[{{Burgasser} {et~al.}(2013){Burgasser}, {Melis}, {Zauderer}, \&
  {Berger}}]{bmzb13}
{Burgasser}, A., {Melis}, C., {Zauderer}, A., \& {Berger}, E. 2013,
  \href{http://dx.doi.org/10.1088/2041-8205/762/1/L3}{ApJL, 762, L3}

\bibitem[{{Burgasser} \& {Putman}(2005)}]{bp05}
{Burgasser}, A., \& {Putman}, M. 2005,
  \href{http://dx.doi.org/10.1086/429788}{ApJ, 626, 486}

\bibitem[{{Burgasser} {et~al.}(2006){Burgasser}, {Geballe}, {Leggett},
  {Kirkpatrick}, \& {Golimowski}}]{bgl+06}
{Burgasser}, A.~J., {Geballe}, T.~R., {Leggett}, S.~K., {Kirkpatrick}, J.~D.,
  \& {Golimowski}, D.~A. 2006, \href{http://dx.doi.org/10.1086/498563}{ApJ,
  637, 1067}

\bibitem[{{Burgasser} {et~al.}(2003{\natexlab{a}}){Burgasser}, {Kirkpatrick},
  {Liebert}, \& {Burrows}}]{bklb03}
{Burgasser}, A.~J., {Kirkpatrick}, J.~D., {Liebert}, J., \& {Burrows}, A.
  2003{\natexlab{a}}, \href{http://dx.doi.org/10.1086/376756}{ApJ, 594, 510}

\bibitem[{{Burgasser} {et~al.}(2003{\natexlab{b}}){Burgasser}, {Kirkpatrick},
  {Reid}, {Brown}, {Miskey}, \& {Gizis}}]{bkr+03}
{Burgasser}, A.~J., {Kirkpatrick}, J.~D., {Reid}, I.~N., {et~al.}
  2003{\natexlab{b}}, \href{http://dx.doi.org/10.1086/346263}{ApJ, 586, 512}

\bibitem[{{Burgasser} {et~al.}(2011){Burgasser}, {Sitarski}, {Gelino},
  {Logsdon}, \& {Perrin}}]{bsg+11}
{Burgasser}, A.~J., {Sitarski}, B.~N., {Gelino}, C.~R., {Logsdon}, S.~E., \&
  {Perrin}, M.~D. 2011,
  \href{http://dx.doi.org/10.1088/0004-637x/739/1/49}{ApJ, 739, 49}

\bibitem[{{Burgasser} {et~al.}(1999){Burgasser}, {Kirkpatrick}, {Brown},
  {Reid}, {Gizis}, {Dahn}, {Monet}, {Beichman}, {Liebert}, {Cutri}, \&
  {Skrutskie}}]{bkb+99}
{Burgasser}, A.~J., {Kirkpatrick}, J.~D., {Brown}, M.~E., {et~al.} 1999,
  \href{http://dx.doi.org/10.1086/312221}{ApJL, 522, L65}

\bibitem[{{Carson} {et~al.}(2011){Carson}, {Marengo}, {Patten}, {Luhman},
  {Sonnett}, {Hora}, {Schuster}, {Allen}, {Fazio}, {Stauffer}, \&
  {Schnupp}}]{cmp+11}
{Carson}, J.~C., {Marengo}, M., {Patten}, B.~M., {et~al.} 2011,
  \href{http://dx.doi.org/10.1088/0004-637x/743/2/141}{ApJ, 743, 141}

\bibitem[{{Chabrier} \& {K\"{u}ker}(2006)}]{ck06}
{Chabrier}, G., \& {K\"{u}ker}, M. 2006,
  \href{http://dx.doi.org/10.1051/0004-6361:20042475}{A\&A, 446, 1027}

\bibitem[{{Condon}(2002)}]{c02}
{Condon}, J.~J. 2002,
  \href{http://adsabs.harvard.edu/cgi-bin/nph-bib_query?bibcode=2002ASPC..278..155C}{ASP~Conf.~Ser.,
  278, 155}

\bibitem[{{Dobler} {et~al.}(2006){Dobler}, {Stix}, \& {Brandenburg}}]{dsb06}
{Dobler}, W., {Stix}, M., \& {Brandenburg}, A. 2006,
  \href{http://dx.doi.org/10.1086/498634}{ApJ, 638, 336}

\bibitem[{{Donati} {et~al.}(2008){Donati}, {Morin}, {Petit}, {Delfosse},
  {Forveille}, {Auri\`{e}re}, {Cabanac}, {Dintrans}, {Fares}, {Gastine},
  {Jardine}, {Ligni\`{e}res}, {Paletou}, {Ramirez Velez}, \&
  {Th\'{e}ado}}]{dmp+08}
{Donati}, J.~F., {Morin}, J., {Petit}, P., {et~al.} 2008,
  \href{http://dx.doi.org/10.1111/j.1365-2966.2008.13799.x}{MNRAS, 390, 545}

\bibitem[{{Dulk}(1985)}]{d85}
{Dulk}, G.~A. 1985,
  \href{http://dx.doi.org/10.1146/annurev.aa.23.090185.001125}{ARA\&A, 23, 169}

\bibitem[{{Durney} {et~al.}(1993){Durney}, {de Young}, \& {Roxburgh}}]{ddyr93}
{Durney}, B.~R., {de Young}, D.~S., \& {Roxburgh}, I.~W. 1993,
  \href{http://dx.doi.org/10.1007/bf00690652}{Solar Physics, 145, 207}

\bibitem[{{Fomalont} {et~al.}(1991){Fomalont}, {Windhorst}, {Kristian}, \&
  {Kellerman}}]{fwkk91}
{Fomalont}, E.~B., {Windhorst}, R.~A., {Kristian}, J.~A., \& {Kellerman}, K.~I.
  1991, \href{http://dx.doi.org/10.1086/115952}{AJ, 102, 1258}

\bibitem[{{Frail} {et~al.}(2012){Frail}, {Kulkarni}, {Ofek}, {Bower}, \&
  {Nakar}}]{fko+12}
{Frail}, D.~A., {Kulkarni}, S.~R., {Ofek}, E.~O., {Bower}, G.~C., \& {Nakar},
  E. 2012, \href{http://dx.doi.org/10.1088/0004-637X/747/1/70}{ApJ, 747, 70}

\bibitem[{{Gizis} {et~al.}(2000){Gizis}, {Monet}, {Reid}, {Kirkpatrick},
  {Liebert}, \& {Williams}}]{gmr+00}
{Gizis}, J., {Monet}, D., {Reid}, N., {et~al.} 2000,
  \href{http://dx.doi.org/10.1086/301456}{AJ, 120, 1085}

\bibitem[{{G\"{u}del} \& {Benz}(1993)}]{gb93}
{G\"{u}del}, M., \& {Benz}, A. 1993,
  \href{http://dx.doi.org/10.1086/186766}{ApJ, 405, L63}

\bibitem[{{Hall}(2002)}]{h02b}
{Hall}, P.~B. 2002, \href{http://dx.doi.org/10.1086/339020}{ApJL, 564, L89}

\bibitem[{{Hallinan} {et~al.}(2006){Hallinan}, {Antonova}, {Doyle}, {Bourke},
  {Brisken}, \& {Golden}}]{had+06}
{Hallinan}, G., {Antonova}, A., {Doyle}, J.~G., {et~al.} 2006,
  \href{http://dx.doi.org/10.1086/508678}{ApJ, 653, 690}

\bibitem[{{Hallinan} {et~al.}(2008){Hallinan}, {Antonova}, {Doyle}, {Bourke},
  {Lane}, \& {Golden}}]{had+08}
---. 2008, \href{http://dx.doi.org/10.1086/590360}{ApJ, 684, 644}

\bibitem[{{Hallinan} {et~al.}(2007){Hallinan}, {Bourke}, {Lane}, {Antonova},
  {Zavala}, {Brisken}, {Boyle}, {Vrba}, {Doyle}, \& {Golden}}]{hbl+07}
{Hallinan}, G., {Bourke}, S., {Lane}, C., {et~al.} 2007,
  \href{http://dx.doi.org/10.1086/519790}{ApJL, 663, L25}

\bibitem[{{McLean} {et~al.}(2012){McLean}, {Berger}, \& {Reiners}}]{mbr12}
{McLean}, M., {Berger}, E., \& {Reiners}, A. 2012,
  \href{http://dx.doi.org/10.1088/0004-637x/746/1/23}{ApJ, 746, 23}

\bibitem[{{McMullin} {et~al.}(2007){McMullin}, {Waters}, {Schiebel}, {Young},
  \& {Golap}}]{thecasa}
{McMullin}, J.~P., {Waters}, B., {Schiebel}, D., {Young}, W., \& {Golap}, K.
  2007,
  \href{http://adsabs.harvard.edu/cgi-bin/nph-bib_query?bibcode=2007ASPC..376..127M}{ASP~Conf.~Ser.,
  376, 127}

\bibitem[{{Mohanty} {et~al.}(2002){Mohanty}, {Basri}, {Shu}, {Allard}, \&
  {Chabrier}}]{mbs+02}
{Mohanty}, S., {Basri}, G., {Shu}, F., {Allard}, F., \& {Chabrier}, G. 2002,
  \href{http://dx.doi.org/10.1086/339911}{ApJ, 571, 469}

\bibitem[{{Morin} {et~al.}(2008){Morin}, {Donati}, {Petit}, {Delfosse},
  {Forveille}, {Albert}, {Auri\`{e}re}, {Cabanac}, {Dintrans}, {Fares},
  {Gastine}, {Jardine}, {Ligni\`{e}res}, {Paletou}, {Ramirez Velez}, \&
  {Th\'{e}ado}}]{mdp+08}
{Morin}, J., {Donati}, J.~F., {Petit}, P., {et~al.} 2008,
  \href{http://dx.doi.org/10.1111/j.1365-2966.2008.13809.x}{MNRAS, 390, 567}

\bibitem[{{Reiners} \& {Basri}(2006)}]{rb06}
{Reiners}, A., \& {Basri}, G. 2006,
  \href{http://dx.doi.org/10.1086/503324}{ApJ, 644, 497}

\bibitem[{{Reiners} \& {Basri}(2007)}]{rb07}
---. 2007, \href{http://dx.doi.org/10.1086/510304}{ApJ, 656, 1121}

\bibitem[{{Route} \& {Wolszczan}(2012)}]{rw12}
{Route}, M., \& {Wolszczan}, A. 2012,
  \href{http://dx.doi.org/10.1088/2041-8205/747/2/l22}{ApJL, 747, L22}

\bibitem[{{Rutledge} {et~al.}(2000){Rutledge}, {Basri}, {Mart\'{\i}n}, \&
  {Bildsten}}]{rbmb00}
{Rutledge}, R.~E., {Basri}, G., {Mart\'{\i}n}, E.~L., \& {Bildsten}, L. 2000,
  \href{http://dx.doi.org/10.1086/312817}{ApJL, 538, L141}

\bibitem[{{Sault} \& {Wieringa}(1994)}]{themfs}
{Sault}, R.~J., \& {Wieringa}, M.~H. 1994,
  \href{http://adsabs.harvard.edu/abs/1994A\%26AS..108..585S}{A\&AS, 108, 585}

\bibitem[{{Semel}(1989)}]{thezdi}
{Semel}, M. 1989,
  \href{http://adsabs.harvard.edu/abs/1989A\%26A...225..456S}{A\&A, 225, 456}

\bibitem[{{Stelzer} {et~al.}(2006){Stelzer}, {Micela}, {Flaccomio},
  {Neuh\"{a}user}, \& {Jayawardhana}}]{smf+06}
{Stelzer}, B., {Micela}, G., {Flaccomio}, E., {Neuh\"{a}user}, R., \&
  {Jayawardhana}, R. 2006,
  \href{http://dx.doi.org/10.1051/0004-6361:20053677}{A\&A, 448, 293}

\bibitem[{{Treumann}(2006)}]{t06}
{Treumann}, R. 2006,
  \href{http://dx.doi.org/10.1007/s00159-006-0001-y}{A\&A~Rev., 13, 229}

\bibitem[{{Vrba} {et~al.}(2004){Vrba}, {Henden}, {Luginbuhl}, {Guetter},
  {Munn}, {Canzian}, {Burgasser}, {Davy Kirkpatrick}, {Fan}, {Geballe},
  {Golimowski}, {Knapp}, {Leggett}, {Schneider}, \& {Brinkmann}}]{vhl+04}
{Vrba}, F.~J., {Henden}, A.~A., {Luginbuhl}, C.~B., {et~al.} 2004,
  \href{http://dx.doi.org/10.1086/383554}{AJ, 127, 2948}

\bibitem[{{West} {et~al.}(2004){West}, {Hawley}, {Walkowicz}, {Covey},
  {Silvestri}, {Raymond}, {Harris}, {Munn}, {McGehee}, {Ivezi\'{c}}, \&
  {Brinkmann}}]{whw+04}
{West}, A.~A., {Hawley}, S.~L., {Walkowicz}, L.~M., {et~al.} 2004,
  \href{http://dx.doi.org/10.1086/421364}{AJ, 128, 426}

\bibitem[{{Wu} \& {Lee}(1979)}]{theecm}
{Wu}, C.~S., \& {Lee}, L.~C. 1979, \href{http://dx.doi.org/10.1086/157120}{ApJ,
  230, 621}

\end{thebibliography}

\end{document}